# Spinning single photons

Yin H. Kan[1,2,†], Sebastian K. H. Andersen[1,†], Fei Ding[1], Shailesh Kumar[1], Chang Y. Zhao[2], Sergey I. Bozhevolnyi[1]*

[1]Center for Nano Optics,

University of Southern Denmark, DK-5230 Odense M, Denmark

[2]Institute of Engineering Thermophysics,

Shanghai Jiao Tong University, Shanghai, 200240, China

[†] These authors contributed equally: Yin H. Kan and Sebastian K. H. Andersen

*Corresponding author's email: seib@mci.sdu.dk

**Single photons carrying spin angular momentum (SAM), i.e., circularly polarized single photons generated typically by subjecting a quantum emitter (QE) to a strong magnetic field at low temperatures[1,2], are at the core of chiral quantum optics enabling non-reciprocal single-photon configurations and deterministic spin-photon interfaces[3]. Here we propose a conceptually new approach to the room-temperature generation of SAM-coded single photons (SSPs) entailing QE non-radiative coupling to surface plasmons that are transformed, by interacting with an optical metasurface, into a collimated stream of SSPs with the designed handedness. We report on the design, fabrication and characterization of SSP sources consisting of dielectric circular nanoridges with azimuthally varying widths deterministically fabricated on a dielectric-protected silver film around a nanodiamond containing a nitrogen-vacancy centre. With properly engineered phases of QE-originated fields scattered by nanoridges, the out-coupled photons feature a well-defined SAM (with the chirality > 0.8) and high directionality (collection efficiency up to 92%).**

Single-photon sources constitute one of the crucial enabling technologies for quantum communications[4-6], quantum computation[7-9], and quantum-enhanced metrology[10-12]. Typical QEs used for realizing single-photon sources feature low emission rates, non-directional emission, and poorly defined polarization properties[13-16], characteristics that prevent QEs from being directly used in quantum technologies[17,18]. By using properly nanostructured environment, i.e., by coupling QEs with nanocavities or nanoantennas, the QE emission rates can be enhanced drastically due to the Purcell effect[19-22]. Many efforts have also been dedicated to obtaining high collection efficiency by directing the QE emission with surrounded nanostructures, including Yagi-Uda antennas[23-26] and bullseye gratings[27-31], demonstrating that highly directional beams can straightforwardly be realized by utilizing different kinds of bullseye structures in forms of concentric ridges. Regarding the single-photon purity and photon indistinguishability, recent reports have proved that these two parameters can simultaneously reach high levels with properly chosen QEs incorporated in semiconductor cavities[32-34]. At the same time, generation of single photons with well-defined polarization properties has rarely been addressed, while the use of (especially) SSPs became very quickly the forefront of research investigations within chiral quantum optics[3] concerned with chiral light-matter interactions in photonic nanostructures that offer fundamentally new functionalities, such as spin-photon interfaces and spin-controlled photon switching[35]. With SSPs at hand it is thus straightforward to realize a chiral waveguide coupler in which the handedness of the incident light determines the propagation direction in the waveguide[3,36]. The direct SSP generation is however a rather complicated and challenging issue: in order to create two non-degenerate circularly polarized QE states one should realize the sufficiently pronounced Zeeman splitting by subjecting QEs, e.g., quantum dots[1] or colour centres in diamond[2], to strong magnetic fields at low temperatures[1,2].

In this Letter, we demonstrate a viable strategy for the room-temperature generation of highly-directional SSPs based on dipolar QEs non-radiatively coupled to surface plasmon-polariton (SPP) modes, which upon interaction with an optical metasurface radiate into free space in the form of a collimated single-photon stream carrying the designed SAM. The proposed configuration represents a hybrid system consisting of a pre-selected nanodiamond (ND) containing a nitrogen vacancy (NV) centre with desired single QE characteristics, which is located in the centre of an optical metasurface composed of hydrogen silsesquioxane (HSQ) circular nanoridges with azimuthally varying widths deterministically fabricated atop a silicon dioxide ($SiO_2$) spacer and silver (Ag) substrate (Fig. 1a). It should be emphasized that, unlike recently introduced quantum metasurfaces interacting with free propagating streams of photons[37,38], the system of nanoridges constituting the metasurface interacts in our case with (non-radiative) SPP modes transforming their fields into unidirectional circularly polarized single photons. Our arrangement (Fig. 1a) can be viewed as a *meta-atom* that, upon illumination with a pump optical wave, would spontaneously emit well-collimated and unidirectional SSPs at room temperature and without magnetic fields applied, a unique property that opens fascinating perspectives in chiral quantum optics, such as, for example, straightforward realization of a single-photon chiral waveguide coupler.

QEs used in our experiments are selected by monitoring their emission when illuminated with a tightly-focused radially-polarized pump laser beam at the wavelength of 532 nm that produces a strong longitudinal electric field component at the focal plane, which is perpendicular to the surface plane. This procedure ensures that the selected QEs have sufficiently large projections of their radiative dipole transitions on the direction perpendicular to the surface. The selected NV-centres, when excited with the pump light, couple efficiently and non-radiatively into cylindrically diverging SPP modes supported by the air-$SiO_2$-Ag interface[35,36], which are subsequently scattered by the circular nanoridges with properly-engineered phases and thus converted into outgoing photons (Fig. 1b). High directionality of emitted photons is secured by matching the nanoridge spacing $P$ to the SPP wavelength calculated for a vacuum wavelength of 665nm, coinciding with the maximum emission of the negatively charged NV state[39]. The polarization state of emitted photons is controlled with the

HSQ metasurface consisting of width-gradient nanoridges that modify the phase of the scattered fields locally by introducing the phase shift $\varphi$ varying with the azimuthal angle $\theta$ (Fig. 1c). By linearly changing the nanoridge width $w$ from 110 nm to 407 nm with the azimuth $\theta$, the relative phase shift $\varphi$ of the scattered electric field is continuously changed within the full $2\pi$ phase range (Fig. 1c). Therefore, the electric fields scattered at two arbitrary points separated by the azimuthal angle $\theta = \pi/2$ (e.g., A and B) are not only orthogonal to each other but also phase-shifted by $\pi/2$, generating thereby circularly-polarized outgoing photons.

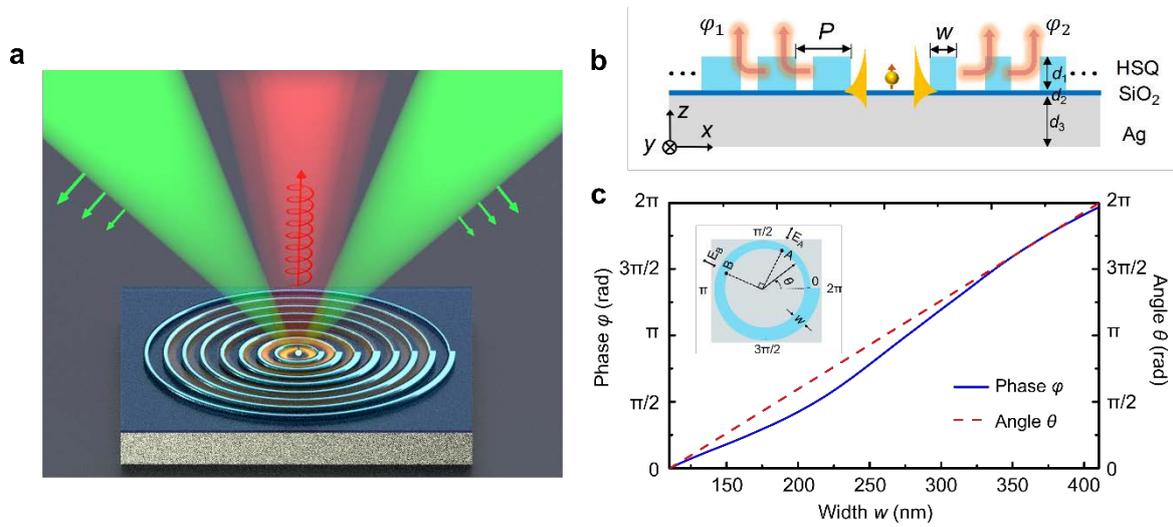

**Figure 1 | Generation of spinning single photons. a,** Schematic of SSP source consisting of a QE with a normal to the surface dipole transition located at the centre of an optical metasurface composed of concentric periodic width-varying dielectric nanoridges atop a thin dielectric spacer, serving for metal protection and reduction of QE quenching, and a metal substrate supporting SPP modes at the QE radiation wavelength. The green cones represent a tightly-focused radially-polarized pump beam that produces a strong longitudinal electric field component at the focal plane. The arrow with a helix in the red beam illustrates an outgoing collimated stream of photons carrying the SAM. **b,** Radial cross section of our configuration showing schematically SPPs, which are excited by an NV-centre, propagating along the surface and out-coupled by scattering off HSQ nanoridges. The metasurface is composed of 150-nm-thick HSQ width-varying nanoridges with the period $P = 550$ nm, a 20-nm-thin SiO$_2$ spacer and a 200-nm-thick Ag substrate. **c,** The simulated phase shift $\varphi$ and azimuthal angle $\theta$ as a function of the nanoridge width $w$, demonstrating that, when the width $w$ varies from 110 nm to 407 nm around the centre, the scattered electric field phase changes continuously within the full $2\pi$ phase range.

The performance of the proposed device is first studied using the three-dimensional (3D) finite-difference time-domain (FDTD) simulation of QE-excited SPP waves scattered by a width-varying system of HSQ circular nanorides (Fig. 2a), in which the refractive index of HSQ is set as 1.41[40] and the QE is treated as an electric dipole oriented normal to the surface in the *z* direction and located at the distance of 50 nm from the SiO$_2$ surface. The QE radiation wavelength of 665nm is chosen to coincide with the maximum emission of the negatively charged NV state[39], resulting in the nanoridge period *P* = 550 nm matching to the SPP wavelength. The radius of the inner ring is then optimized to be 585 nm to ensure the highest quantum efficiency at the design wavelength of 665 nm along with the reasonably high Purcell factor of $\cong$ 4.0 (Supplementary Section 1). For the considered configuration, the quantum efficiency, $\eta_{QE}$, defined as the ratio between the radiative and total decay rates[41] is found to be $\eta_{QE} \cong 0.46$, a somewhat low value due a relatively low refractive index of HSQ nanoridges that can however be improved by using high-index dielectrics, e.g., titanium dioxide (TiO$_2$)[27], up to 0.8. The intensity of the resulting out-coupled QE radiation is found concentrated near the perpendicular to the surface plane direction (Fig. 2b) with the divergence angle of only $\cong 4°$, as determined at the full-width-at-half-maximum (FWHM) of intensity distribution. The fraction of radiative light collected by the first objective with a numerical aperture (NA) of 0.9 is calculated to be $\cong 0.92$ (i.e., the collection efficiency $\eta_{ce}$ of 92%).

We evaluate the degree of circular polarization in the far field by the normalized Stokes parameter defined as $P_c = (I_R - I_L)/(I_R + I_L)$, where $I_R(I_L)$ denotes the intensity for the right-hand (left-hand) circular polarization ($P_c = 1, 0, -1$ represents the ideal right-hand circular, linear polarization, and left-hand circular polarizations, respectively). It is seen that, for the considered design (Fig. 2a), the right-hand circular polarization is observed at the central area of the angular distribution, i.e., of the Fourier plane (Fig. 2c), i.e., precisely where the maximum intensity is located (cf. Fig. 2b). This matching becomes even more transparent when both the far-field intensity and polarization angular distributions are 3D displayed with the height being proportional to the far-field intensity and the colour to the field polarization state (Fig. 2d). It is also seen that some radiation would come out with the left-hand circular polarization, indicating that the design can further be perfected. It is clear that, while the azimuthally varying width of nanoridges serves the purpose of controlling the *phase* of out-coupled radiation, this gradient introduces azimuthal variations in the radiation *amplitude*, which is an undesirable (though not strong) effect causing asymmetry in the out-coupled radiation pattern and imbalance in the resulted polarization. The finite numbers of out-coupling nanoridges and simplified design, in which we use for simplicity the linear width-gradient (Fig. 1c), might also contribute to deviations from the ideal performance. Nevertheless, the full 3D simulations of the considered configuration demonstrate that the circularly polarized QE emission can be realized with an integrated metal-dielectric hybrid system composed of a width-gradient nanoridge metasurface interacting with SPP waves excited by a well-aligned QE.

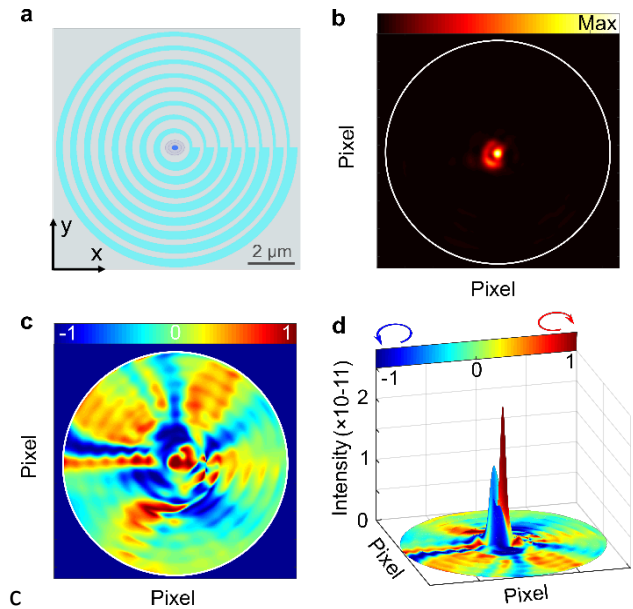

**Figure 2 | 3D simulation of SSP generation. a**, Schematic top view of the width-gradient nanoridge metasurface a QE. **b**, Far-field angular intensity distribution (in the Fourier plane) of the out-coupled QE radiation, featuring a very bright spot at the centre, i.e., near the perpendicular to the surface plane direction, with the divergence angle of only 4.01°. **c**, Far-field polarization state distribution (in the Fourier plane) of the out-coupled QE radiation with the red colour corresponding to the right-hand circular polarization and the blue to left-hand one. **d**, 3D representation of the superimposed beam intensity and polarization far-field distributions, with the height reflecting the light intensity and the colour indicating the polarization. The dominant peak is red-coloured, demonstrating that the majority of photons carry the designed SAM, i.e., possess the right-hand circular polarization.

The experimental demonstration of SSP generation involves several technological steps requiring proper adjustment of nanofabrication parameters. To facilitate the fine-tuning, we first conducted the essential experimental steps using ~ 100-nm-diameter NDs containing on average ~ 400 NV centres per ND (Adamas Nanotechnologies). These NDs can easily be located with the dark-field microscopy producing very bright spots (Supplementary Section 2). After spin-coating NDs on the 20-nm-thin $SiO_2$ spacer atop Ag substrate, the relative position of a selected ND with respect to prefabricated gold aligning markers is determined using the corresponding dark-field microscopy image. The HSQ width-gradient circular nanoridges (Figs. 3a and 3e) are subsequently fabricated around the selected NDs with standard electron-beam lithography using the aligning markers (Supplementary Section 2). The fabricated meta-atom arrangements with the selected ND surrounded by nanoridges are then excited with a tightly-focused radially-polarized 10-$\mu$W-pump cw-laser beam at the wavelength of 532 nm that produces a strong longitudinal electric field component at the focal plane, which is perpendicular to the surface plane (Supplementary Section 3.1). This optical pump configuration ensures that thus excited NV centres (in the selected NDs) have sufficiently large projections of their radiative dipole transitions on the direction perpendicular to the surface, making thereby the experimental conditions to be similar to those used in our simulations (Fig. 2).

The far-field distributions of intensity and polarization state of the emitted radiation can be determined by a set of intensity measurements, giving the four Stokes parameters with $S_0$ as the total intensity of the emission, $S_1$ - the intensity of linear horizontal or vertical polarization, $S_2$ - the intensity of linear $+45°$ or $-45°$ polarization, and $S_3$ - the intensity of right- or left-hand circular polarization. These Stokes parameters are determined by the intensities measured with a properly-oriented quarter waveplate and linear polarizer mounted on rotation stages (Supplementary Section 3.2). The far-field intensity distributions for the both realized configurations (Figs. 3b and 3f) demonstrate well-collimated out-coupled beams with the FWHM divergence of $\cong 6.1°$, which is understandably larger than the simulated value of $\cong 4°$ because of spatial spread in NV centres inside NDs. We evaluate the chirality of emitted photons using a modified $S_3$-parameter that is corrected for unpolarized light and defined as $P_c = S_3'/\sqrt{(S_1')^2 + (S_2')^2 + (S_3')^2}$, in which $S_1'$, $S_2'$, and $S_3'$ are the Stokes parameters normalized to the corresponding total intensity ($S_0$) obtained in each measurement. The far-field chirality distributions indicate that the majority of photons (deduced from the corresponding intensity distributions) carry the designed SAM, i.e., possess the right- (Fig. 3c) or left-hand (Fig. 3g) circular polarization with $P_c \geq 0.8$. The intended intensity-chirality matching becomes even more transparent when both the far-field intensity and polarization angular distributions are 3D displayed with the height being proportional to the far-field intensity and the colour to the field polarization state (Figs. 3d and 3h). Overall, these experiments with multi-photon emitters demonstrate the feasibility of our approach for generation of photons carrying the designed SAM.

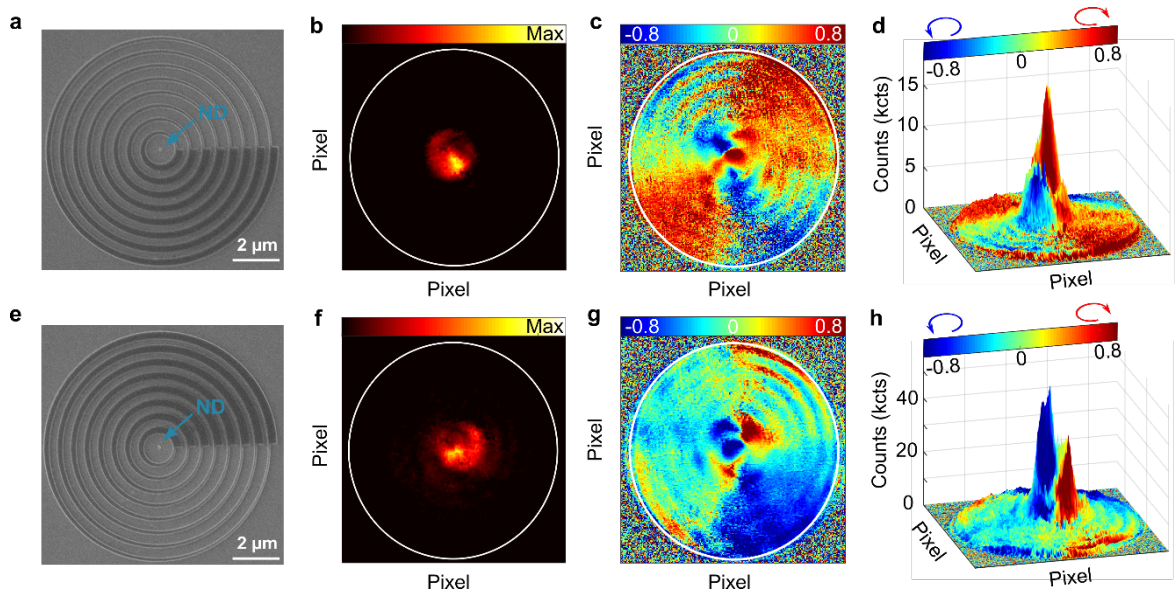

**Figure 3 | Experimental demonstration of SAM-coded photon generation. a**, SEM image of circular width-gradient nanoridges for generation of right-hand circularly polarized photons. The selected ND with many NV centres is situated in the centre of circular nanoridges. **b**, The measured (in the microscope objective back-focal plane) far-field intensity distribution of outgoing photons. The white circle indicates the extent of objective NA (NA = 0.9). **c**, Far-field polarization state distribution (in the Fourier plane) of the out-coupled radiation with the red colour corresponding to the right-hand circular polarization and the blue to left-hand one. **d**, 3D representation of the superimposed beam intensity and polarization far-field distributions, with the height reflecting the light intensity and the colour indicating the polarization. The dominant peak is red-coloured, demonstrating that the majority of photons carry the designed SAM, i.e.,

possess the right-hand circular polarization. **e-h**, Experimental results similar to those shown in **a-d** obtained with the width-gradient nanoridges of opposite chirality needed for the generation of left-hand circularly polarized photons.

The demonstration of SSP generation requires the usage of NDs containing single colour centres. For this purpose, we selected ~ 30-nm-diamter NDs with a large portion containing single NV centres (MicroDiamant). Due to the small size these NDs have very low scattering cross sections (roughly 3 order of magnitudes smaller than that previously used with many NV centres) and are very problematic to localize on the dark-field microscopy images. In these experiments, the single-vacancy ND positions are determined using fluorescence images that are obtained using considerably higher laser powers (~ 200 $\mu$W) under otherwise the same conditions than in the experiments with many NV centres (Fig. 4a). The relatively weak single-photon emission from the selected ND leads to a relatively poor signal-to-noise ratio, especially in the final configuration with the fluorescence from the fabricated HSQ metasurface being clearly visible (Fig. 4e), a detrimental circumstance that serves however the purpose of enabling to assess the accuracy of the positioning of HSQ nanoridges with respect to the ND.

The single-vacancy ND used for demonstrating the SSP generation is systematically characterized before and after the fabrication the metasurface width-gradient circular nanoridges, i.e., in the uncoupled and coupled (to the metasurface) configurations. A band-pass filter with the spectral range of $650 \pm 20$ nm that includes the strongest emission of the single NV centre (Fig. 4i) is used to characterize the far-field intensity and polarization state distributions (Supplementary Sections 3.2 and 3.3). As anticipated, before the metasurface fabrication, the measured far-field fluorescence intensity is homogenously distributed (Fig. 4b), and no circular polarization is observable (Fig. 4c). After the fabrication, the intensity distribution demonstrates a well-collimated beam concentrated in the centre with the angular FWHM of $\cong 4.7°$ (Fig. 4f), which is slightly larger than the simulated value of $\cong 4°$ (Fig. 2b) but smaller than that observed ($\cong 6.1°$) with many NV centres (Fig. 3b). These differences might be considered as indications that the NV centre is located slightly off the metasurface centre but not as much as some NV centres contained in the large (~ 100-nm-diameter) NDs used in the experiments with many NV centres. The corresponding polarization state distribution (Fig. 4g) reveals that the main emission detected is characterized by the right-hand circular polarization ($P_c > 0.8$ in the centre). The intended intensity-chirality matching becomes even more transparent when both the far-field intensity and polarization angular distributions are 3D displayed (Figs. 4j), although the noise background blurs the resulting image (cf. Figs. 3d and 4j).

Fitting a 3-level model to the second-order correlation data confirms that that the selected NV centre remains being a single-photon source throughout the experiment, although the second-order correlation minimum deteriorates from $g^{(2)}(0) \cong 0.17$ (Fig. 4d) to $g^{(2)}(0) \cong 0.27$ (Fig. 4h) measured, correspondingly, before and after the metasurface fabrication. As mentioned above the decrease in the signal-to-noise ratio is associated with the occurrence of the fluorescence from the fabricated HSQ metasurface. In general, the NV emission becomes significantly larger after the metasurface fabrication, primarily due to the redirecting of SPP waves (excited by the pumped NV centre) towards the detection by scattering off the nanoridges. Thus, the saturated photon rate, $R_\infty = \eta QE/\tau$, measured without the band-pass filter, increases by a factor $R_\infty/R_{\infty,0} = 2.74$, from $R_{\infty,0} = 104$ kcps to $R_\infty = 284$ kcps after introducing the width-gradient nanoridges, while the saturation power remained nearly unchanged $P_{s,0} = 320$ μW before and $P_s = 391$ μW after the metasurface fabrication (Fig. 4k). A slight increase in the latter might simply be related to the additional contribution from HSQ fluorescence. The increase in the saturated photon rate is in good agreement with the numerical simulations predicting the ratio $R_\infty/R_{\infty,0} = 3.29$. Considering the fact that only a minor lifetime enhancement of $\tau_0/\tau = 1.4$ is expected from our simulations matching the experimentally observed enhancement $\tau_0/\tau = 1.2 - 1.5$ (Supplementary Section 3.4), we infer that the increased photon rate is primarily due to an improvement in the quantum efficiency that results from the SPP waves being scattered to free space. Indeed, our modelling predicts a nearly double increase in the quantum efficiency, from $\eta_{QE0} \cong 0.24$ to $\eta_{QE} \cong 0.46$, after the introduction of metasurface, while the collection efficiency increases only slightly, from $\eta_0 = 0.75$ to $\eta = 0.92$ (when considering a relatively large NA of 0.9).

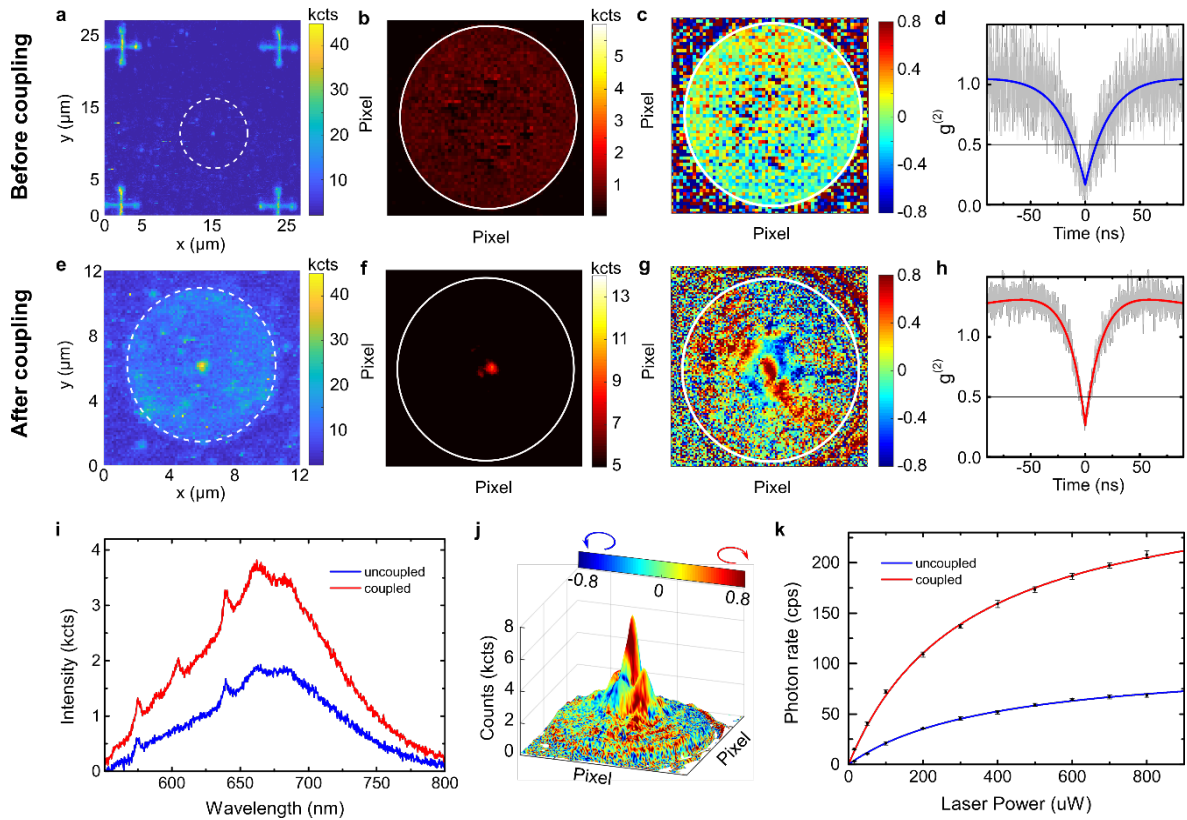

**Figure 4 | Experimental demonstration of SSP generation. a**, The fluorescence image featuring the selected ND – a bright spot in the centre of a white-dashed circle indicating the area to be occupied by metasurface nanoridges. Aligning markers (crosses) are also seen due to the gold fluorescence. **b, f**, The measured (in the microscope objective back-focal plane) far-field emission intensity distributions of uncoupled, **b**, and coupled, **f**, NV centre. The white circle indicates the extent of objective NA (NA = 0.9). **c, g**, Far-field polarization state distributions (in the Fourier plane) of the emission of uncoupled, **c**, and coupled, **g**, NV centre, with the red colour corresponding to the right-hand circular polarization and the blue to left-hand one. **d, h**, The second-order correlation function g$^{(2)}$(t) measured before, **d**, and after, **h**, the

metasurface fabrication. **e,** The fluorescence image featuring the selected ND as a bright spot in the centre of a weakly fluorescing circular area representing the HSQ metasurface fabricated with the same parameters as that shown in Fig. 3a. The fact that the ND-related bright spot is located exactly at the centre of the metasurface-related circle confirms the accuracy of the positioning of HSQ nanoridges with respect to the ND. **i,** Emission spectra of the selected ND before (blue) and after (red) the metasurface fabrication. **j,** 3D representation of the superimposed beam intensity and polarization far-field distributions, with the height reflecting the light intensity and the colour indicating the polarization. The dominant peak is red-coloured, demonstrating that the majority of photons carry the designed SAM, i.e., possess the right-hand circular polarization. **k,** Emission saturation dependencies on the pump laser power before (blue) and after (red) the metasurface fabrication.

The proposed and demonstrated general approach to the implementation of complete control over scattering wavefronts emerging from of a single distributed (in the surface plane) SPP quantum generated by an excited QE opens the doorway to engineering integrated photon sources capable of generating SAM-coded single photons with controllable far-field distributions at room-temperature. This approach can straightforwardly be extended to generation of single photons carrying also orbital angular momenta and composing, in general, vector beams. Further improvements of the experimental realizations include the use of high-refractive index nanoridges, e.g., made of titanium dioxide ($TiO_2$)[27], and gap-plasmon resonators[41,42] in order to boost up the quantum efficiency (up to 0.8) and the Purcell enhancement of the emission rates (by $10^4$ times), respectively. Moreover, the purity, indistinguishability can be improved in the future by using other QE configurations (e.g., quantum dots) and semiconductor materials (e.g., AlGaAs, InGaAs)[32-34]. While the current grating design is optimized for a $2\pi$ phase variation with the azimuthal angle, it is further desirable to maintain a constant scattering amplitude. We envision the control and management of the scattering amplitude by varying the height of nanoridges, e.g., by making use of gray-scale electron beam lithography[43]. The access to engineering SAM-coded single-photon states may find numerous applications in quantum key distribution, interaction with chiral QE transitions and polarization filtering of laser light in resonant fluorescence experiments[32]. Overall, the developed room-temperature SSP generation approach opens new fascinating perspectives within integrated optical quantum technologies, in general, and chiral quantum optics[3], in particular.

**Acknowledgements**

The authors gratefully acknowledge financial support from the European Research Council, Grant 341054 (PLAQNAP). S.I.B. acknowledges the support from the Villum Kann Rasmussen Foundation (Award in Technical and Natural Sciences 2019). Y.H.K. and C.Y.C. acknowledge the support from the National Natural Science Foundation of China (Grants No. 51636004).

**Author contributions**

Y.H.K. and S.K.H.A contributed equally to this work. S.I.B. conceived the configuration geometry. Y.H.K. with assistance from F.D. performed theoretical modelling. Y.H.K. and S.K.H.A with assistance from F.D. and S. K. carried out the experiments. Y.H.K., S.K.H.A, F.D. and S.I.B. analysed the data. S.I.B., F.D, and C.Y.C supervised the project. Y.H.K. wrote the manuscript with contributions from all other authors.

## SUPPLEMENTARY INFORMATION

### 1. Numerical Modelling

Numerical simulations are performed with commercial FDTD software package (FDTD Solutions, Lumerical Solutions). The refractive indexes of Ag and SiO2 are taken from the handbook of optical constants of solids[s1]. We set the refractive index of HSQ as 1.41. The source is an electric dipole direction perpendicular to SiO2 layer surface. The simulation is performed for a vacuum wavelength of $\lambda_0 = 665$ nm to coincide with the peak emission of the negative charge state of the NV centre. Our design of the circularly polarized single-photon device starts with estimating the period of the bullseye gratings by 2D FDTD simulation, to ensure directional emission normal to the surface. Surface plasmon polaritons (SPPs) scatter to free space radiation, according to the phase-matching condition in equation (1)[s2].

$$k_{spp} = k_\sigma \pm nG \qquad n = \pm 1,2,3, \ldots \quad (1)$$

$G = 2\pi/\Lambda$ being the grating vector with period $\Lambda$, $k_\sigma = k_0 \sin\theta$ being the in-plane wavevector of free radiation with propagation constant $k_0 = 2\pi/\lambda_0$, propagating along emission angle $\theta$, normal to the surface. $k_{spp} = k_0 N_{eff}^G$ is the SPP propagation constant with effective index $N_{eff}^G$. For the dielectric grating we describe $N_{eff}^G$, weighted by grating fill factor $\chi$ as:

$$N_{eff}^G = (1-\chi)N_{eff1} + \chi N_{eff2}. \quad (2)$$

We calculated the period of SPPs and effective indexes for the scenarios without and with HSQ layer, as shown in Fig. S1. The results show that $N_{eff1} = 1.101$ being the SPPs effective index for Air-SiO2-Ag regime and $N_{eff2} = 1.446$ for HSQ-SiO2-Ag regime. For $\chi = 0.3$ we find $N_{eff}^G = 1.204$. Corresponding for out of plane scattering we find the period $\Lambda = \frac{\lambda_0}{N_{eff}^G} \approx 550$ nm. Figure S1(c) is the far-field intensity distribution of out-coupling emission for circular nanoridges with $\Lambda = 550$ nm, $\chi = 0.3$, which demonstrate under this period the SPPs could coupling into free space directional beam.

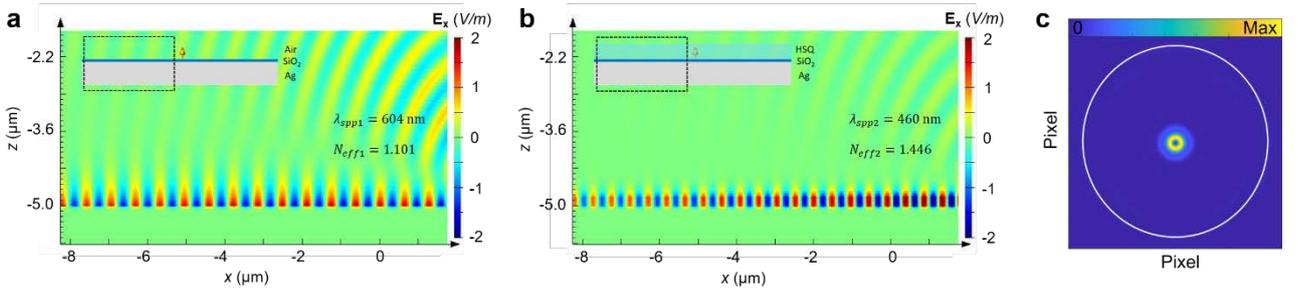

Fig. S1: The wavelength and effective indexes of SPPs propagating along the interface between SiO2 and Ag, **a**, without HSQ layer, **b**, with HSQ layer. The thickness of Ag substrate is 200 nm, followed by 20 nm SiO2. The SPPs are excited by an electric dipole situating 50 nm above the SiO2 layer surface. The dash line rectangles in inset indicate the displayed area of the simulation domain for obtaining the distribution profiles of the electric field component $E_x$. **c**, Far-field intensity distribution of out-coupling emission for circular nanoridges with $\Lambda = 550$ nm, $\chi = 0.3$.

After knowing the period of the gratings, we analyse the relationship between the phase of the SPPs scattered light as a function of the grating width. In the calculation, we vary the width $w$ of the HSQ ridges from 100 nm to 450 nm, while monitoring the phase of $\mathbf{E}_x$, above the grating (Fig S2 a,b). The results in main text Fig. 1 c shows that with the width changing from 110 nm to 407 nm, the phase of $\mathbf{E}_x$ could cover 0 to $2\pi$.

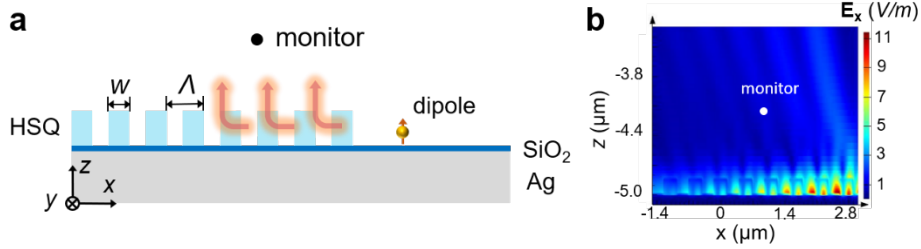

Fig. S2: Numerical calculation of the relationship between the phase of emission light with the HSQ grating widths. **a**, Schematic of 2D simulation. **b**, Electric intensity profile with the grating width $w = 300$ nm.

The inner radius is optimized to obtain the optimal Purcell factor $P_f = \gamma_{sp}/\gamma_{sp}^0$ and quantum efficiency $QE = \gamma_r/\gamma_{sp}$, where $\gamma_{sp}^0$ is the spontaneous emission rate in free space, $\gamma_{sp}$ is total spontaneous emission rate in the antenna structure, and $\gamma_r$ is the radiative rate of the emitter. In our simulation, $QE$ is calculated by the ratio of the radiated power measurable in the far field to the total power injected by a dipole source. We set the inner radius $r_0 = 585$ nm ($P_f = 4.03, QE = 0.46$).

In the simulation the degree of circular polarization is defined as $P_c = (I_R - I_L)/(I_R + I_L)$, where $I_R$ and $I_L$ are intensities of the left and right handed circular polarization components. The far field projection function is used to project the near-field electric field into linear far-field vector field components[s3]. By converting the linear polarization component to their equivalent circular polarization components, we can express all the polarization light into circular polarization components. In the far field, intensity is proportional to abs(E)^2.

## 2. Sample fabrication

The samples are fabricated by ohmic evaporation of 3 nm Ti, 200 nm Ag, 3 nm Ti topped by RF-sputtering of 20 nm SiO$_2$ on a Si wafer, at deposition rates of 0.1 Å/s, 1 Å/s, 0.1 Å/s, and 1.5 Å/s and $5 \times 10^{-6}$ mbar chamber pressure. The Ti films act as adhesion layers. PMMA A2 is subsequently spincoated at 1580 rpm (45 s) on the sample and prebaked at 180°C for 2 min before patterning alighmarkers by a 30-kV electron beam lithography system (JEOL-6490). After development (1:3 MIBK-to-IPA for 35s followed by 1min rinse in IPA), a 35 nm Au layer is deposited by ohmic evaporation at 1 Å/s and $5 \times 10^{-6}$ mbar chamber pressure. Lift-off is performed in acetone for 6 h to reveal the alignment markers. For experiments with multiple NV-centres, 100 nm NDs containing ~400 NV-centres (Adamas technology) are first spincoated on the samples. The relative positions of the NDs in the coordinate frame of the alignment marks are determined by a dark-field microscope image (Fig. S3). A coordinates system is initially defined by selecting the lower left marker as the point of origin. Selecting the lower right marker, defines the x-axis and physical dimensions, by scaling pixel length to the known distance between selected markers. A well-defined xy-coordinate is confirmed by plotting the estimated position of the top markers. The ND coordinates is finally obtained by fitting two Gaussians to the diffraction-limited spot of the ND.

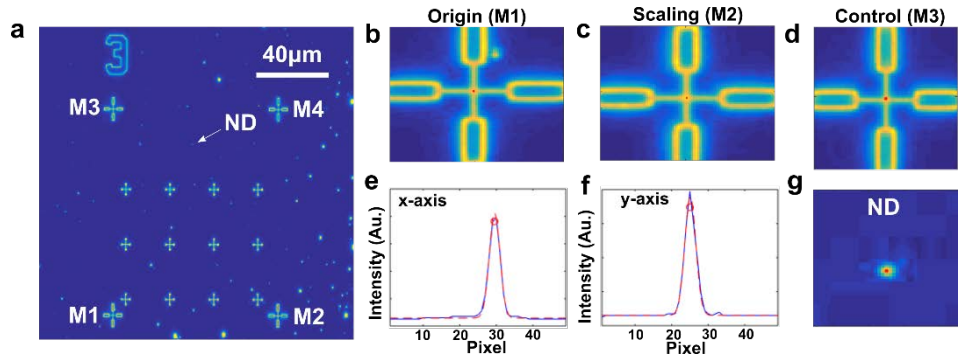

Fig. S3 Alignment procedure for determining the position of a ND. **a**, Starting point is dark-field image in grayscale, showing four alignment marks M1-M4 and the target ND. **b**, A coordinate system is defined by setting the centre of M1 to origin, indicated by red spot. **c**, A Horizontal coordinate axis and scaling is defined by selecting the centre of M2. **d**, A control of the defined coordinate system is done, by plotting expected positions of M3 and M4. **e**, **f** The coordinate position of the ND is determined fitting two gaussians fitting (red) to the diffraction-limited spot of the ND (blue). **g**, Red spot the final estimated position of the ND.

An HSQ layer was subsequently spincoated at 1000 rpm (60 s) and prebaked at 160℃ for 2 min. The gradient-width bullseye gratings are patterned around the ND by electron beam lithography by aligning to the makers, followed by development using 25% TMAH (Tetramethylammonium hydroxide) for 4 min. Single photon experiments were conducted using natural ND 0-0.05 μm GAF (Microdiamant). As so small ND's were not visible in dark-field images and few ND's contain NV-centres, the NV position was determined based on fluorescence maps instead using the same image treatment as for ND with multiple NV-centres.

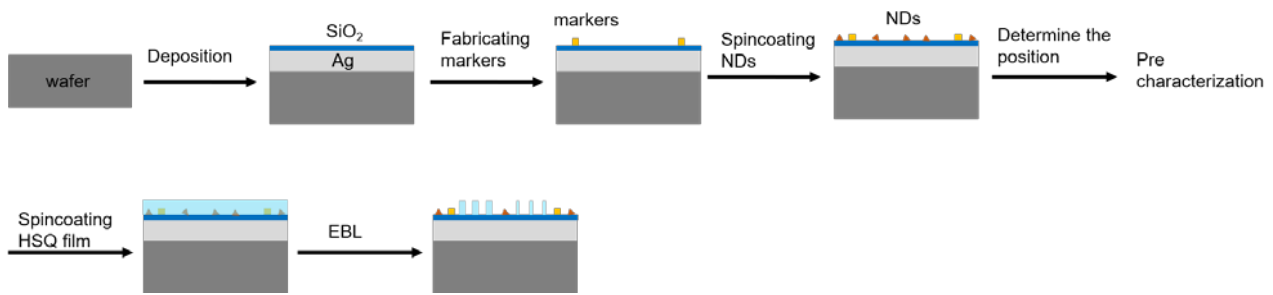

Fig. S4: Process for fabrication spinning photon sources.

### 3. Characterizations of the samples

## 3.1 Experimental setup

In Fig. S5, we present the schematic of the optical set-up that we used for characterizing our NDs, as well as NDs-nanoridges coupled systems. The pump laser is a linearly polarized 532 nm continuous wave (Crystal laser) or pulsed (LDH-P-FA-530L; Pico Quant) chosen by a flip mirror (FM). A half-wave plate ($\lambda/2$) in combination with a Glan-Taylor polarizing beam splitter (PBS) allow for power control and lock the axis of linear polarization. A liquid crystal display (ARCoptix RPC; ARCoptix) convert the linear polarized light to a radially polarized beam. The beam is focused onto the sample using a x100 0.9 NA objective (MPLFLN x100; Olympus). Fluorescence collected by same objective is filtered from the laser light, by a set of dichroic mirrors (DM) (FF535-SDi01/FF552-Di02; Semrock) and a long pass filter (LPF) (FELH0550; Thorlabs). Recording the fluorescence photon rate with an avalanche photo diode (APD1) ($\tau - SPAD$, Pico quant), while scanning the sample, using a piezo-stage, allowed for locating NV-centers by the recording of fluorescence maps. The spectrum of the NV-center was measured with a spectrometer equipped with a CCD camera (Ultra 888 USB3 –BV, Andor). 2. Order correlation measurements was recorded by histogramming the timing delay between photon detection events between APD1 and APD2 in a start-stop configuration, using an electronic timing box (Picharp-300; Pico quant). For determining the Stokes parameters, a broadband quarter wave plate ($\lambda/4$) (SAQWP05M; Thorlabs) mounted on a motorized rotation stage is flipped into the optical path together with a polarizer (LPVIS100-MP2; Thorlabs). A flip mirror projects the Fourier plane onto a CCD camera (Orcad4LT; Hamamatsu). Fourier plane images of $S_3$ was obtained by setting the fast-axis of the quarter-wave plate to respectively $\pm 45°$ wrt. to the polarization axis. For measuring the S1 and S2 stokes-parameters, the quarterwave plate is removed and the polarizer is mounted on the motorstage, and measurements performed according to section 3.2.

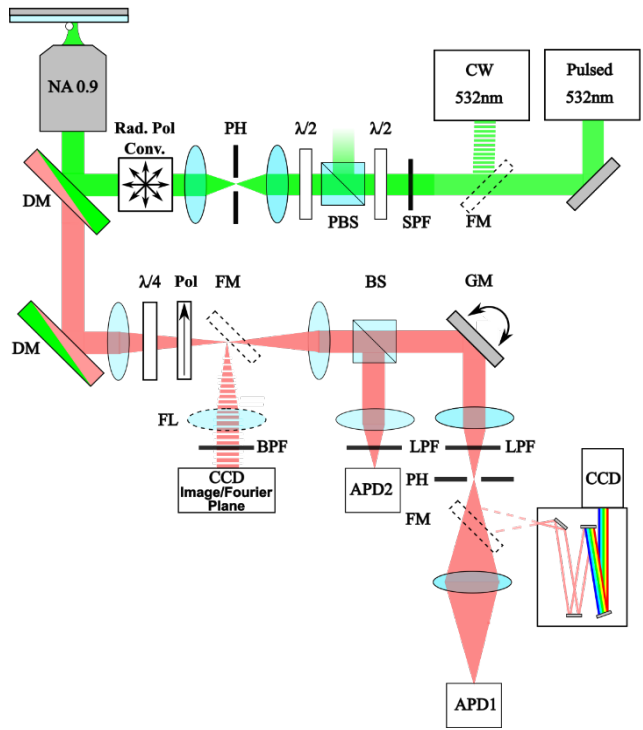

Fig. S5: Experimental setup for characterization of emission properties of quantum emitters. FM: Flip mirror, SPF: Short pass filter, $\lambda/2$: Half-wave plate, PBS: Polarizing beam splitter, PH: Pinhole, Rad.Pol.Conv: Linear-to-radial polarization converter. DM: Diachroic mirror. $\lambda/4$: Quarter-waveplate. Pol: Analyzer. FL: Flip lens. BPF: Band pass filter (Thorlabs 650 nm ±20 nm, 676nm ±20 nm, and 700nm ±20 nm). BS: 50:50 beam splitter. GM: Galvanometric mirror, LPF: Long pass filter. APD: Avalanche photo diode.

### 3.2 Measurement of Stokes parameters

The polarization state of light may be determined by a set of intensity measurements, giving the four Stokes parameters.

$$S_0 = I_{tot} = I_{pol} + I_{unpol}, \tag{3}$$

$$S_1 = I_H - I_V, \tag{4}$$

$$S_2 = I_{+45} - I_{-45}, \tag{5}$$

$$S_3 = I_{RC} - I_{LC}. \tag{6}$$

Where $S_0$ is the total intensity of the emission (polarized+unpolarized), $S_1$ is the intensity of linear horizontal ($I_H$) or vertical polarization ($I_V$), $S_2$ is the intensity of linear $+45°$ ($I_{+45}$) or $-45°$ ($I_{-45}$) polarization, and $S_3$ is the intensity of right ($I_{RC}$) or left ($I_{LC}$) circular polarization.

It can be demonstrated that the intensity measured after retarder plate with retardation phase $\varphi$ (fast axis along x) and linear polarizer at angle $\theta$, $wrt.\ x - axis$, is given by:

$$I(\theta, \varphi) = \tfrac{1}{2}[S_0 + S_1 \cos(2\theta) + S_2 \cos(\varphi)\sin(2\theta) + S_3 \sin(\varphi)\sin(2\theta)]. \tag{7}$$

With only polarizer in the optical path ($\varphi = 0$), from Eq. (7), we can obtain:

$$S_1 = I(0,0) - I(90°, 0), \tag{8}$$

$$S_2 = I(45°, 0) - I(-45°, 0), \tag{9}$$

With both polarizer and quarter wave plate in optical path ($\varphi = 90$),

$$S_3 = I(-45°, 90°) - I(-45°, 90°), \tag{10}$$

$$S_0 = I(-45°, 90°) + I(-45°, 90°). \tag{11}$$

The total intensity $S_0$ is

$$S_0 = I_{pol} + I_{unpol} = \sqrt{S_1^2 + S_2^2 + S_3^2} + I_{unpol}. \tag{12}$$

To characterize the polarization of emission light, we define the degree of circular polarization $P_c$ as,

$$P_c = \frac{S_3'}{\sqrt{(S_1')^2 + (S_2')^2 + (S_3')^2}}. \tag{13}$$

Here, $S_1'$, $S_2'$, and $S_3'$ are normalized to the corresponding total intensity in each measurement. Fig. S6 presents the normalized $S_1'$, $S_2'$, $S_3'$ and $P_c$ for the samples with many NV centres coupling with right-hand width-gradient circular nanoridges Fig. 6 (a)-(d) and left-hand width-gradient circular nanoridges Fig. 6 (e)-(h) and.

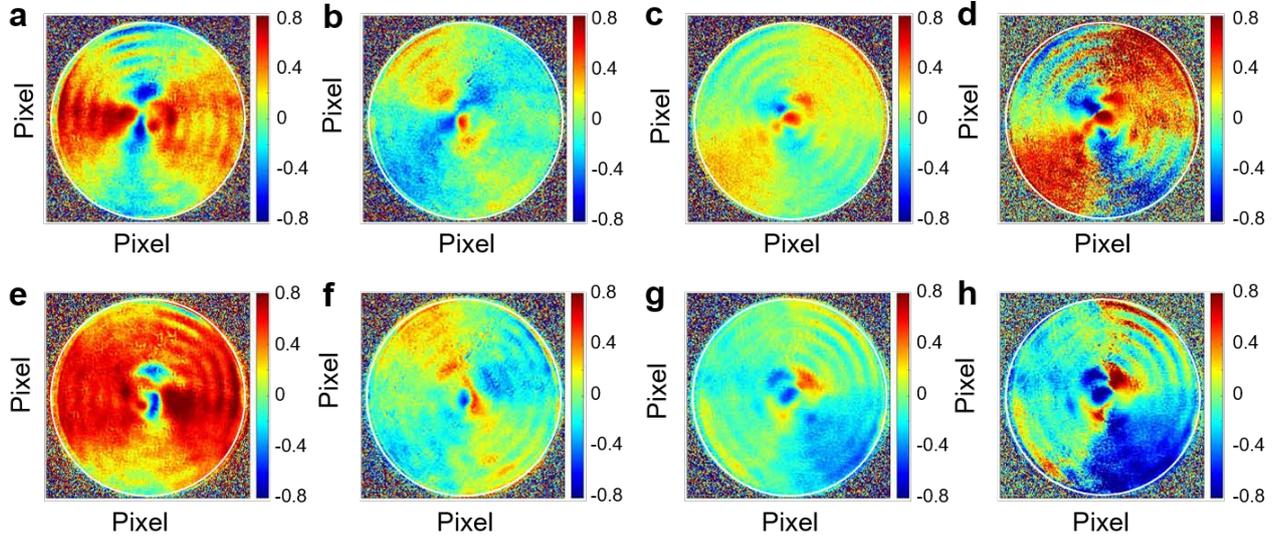

Fig. S6 Normalized Stokes parameters for many NV-centres samples. **a**, $S_1'$, **b**, $S_2'$, **c**, $S_3'$, and **d**, $P_c$ for right-hand width-gradient circular nanoridges. **e**, $S_{1_1}'$, **f**, $S_2'$, **g**, $S_3'$, and **h**, $P_c$ for left-hand width-gradient circular nanoridges.

### 3.3 Wavelength-dependence of $S_0$ and $S_3$

In order to investigate the wavelength dependent performance of the fabricated width-gradient circular nanoridges the $S_0$ and $S_3$ Stokes-parameter were measured in the Fourier plane, while filtering the fluorescence through various band-pass filters covering the spectral range of the NV-center. Figure S7 show the fluorescence spectrum of the examined NV-centers situated in the width-gradient circular nanoridges and indicate the spectral range of the respective band-pass filters. Band pass filters with center wavelength 600nm, 650nm, 700nm and 750nm (FB#; Thorlabs) have a 40nm transmission band-width, while the band-pass filter with center wavelength 676nm (Edmund Optics) have a 29nm transmission band-width.

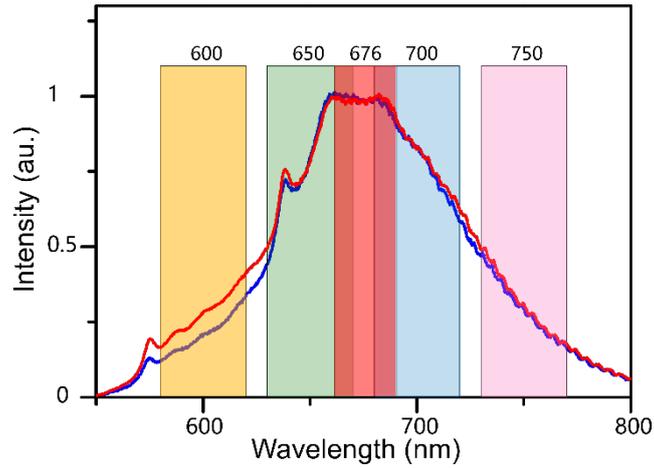

Fig. S7  Fluorescence spectrum of NV-centers situated in width-gradient circular nanoridges, further examined of wavelength dependent response in Fig. S8 top panel (blue) and lower panel (red). Transparent boxes indicate the transmission band of band-pass filters used for filtering with a center wavelength indicated above the respective boxes.

The spectrally filtered Fourier images of the $S_0$ and $S_3$ are presented in Fig. S8. For filters selecting the spectral range 630-720nm, we observe beaming ($S_0$) correlated with circular polarized emission ($S_3$), for both the examined antenna systems. The wavelength dependent performance is in good agreement with the numerical design, optimized for the wavelength 665nm, situated near the center of the spectral range.

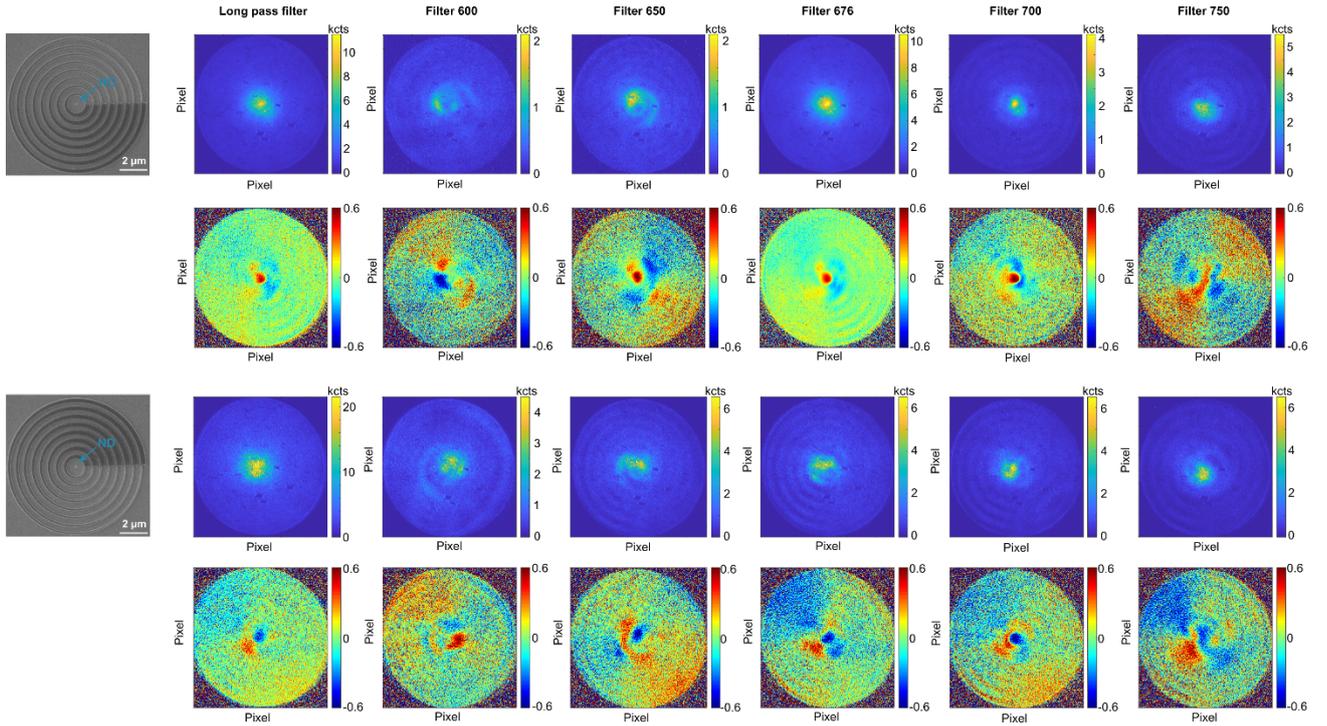

Fig. S8 Far-field intensity and normalized Stokes parameters for many NV-centres samples with different band pass filters.

Fig. S9 shows the measured $S'_1$, $S'_2$, $S'_3$ and $P_c$ for single NV centre coupling with right-hand width-gradient circular nanoridges.

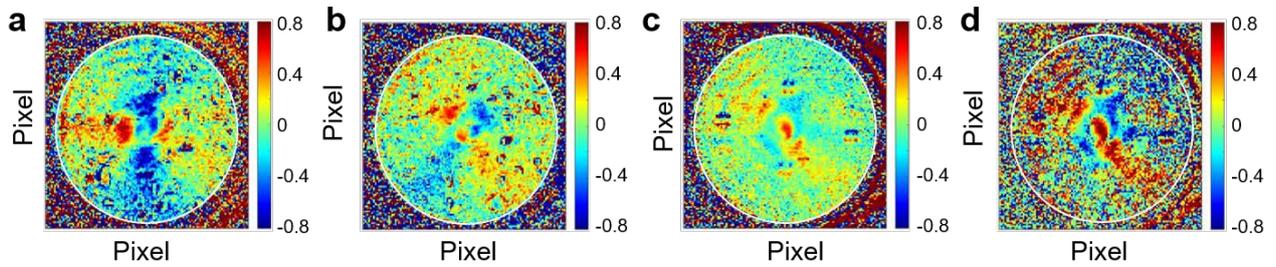

Fig. S9 Normalized Stokes parameters for single NV-centre sample. **a**, $S'_1$, **b**, $S'_2$, **c**, $S'_3$, and **d**, $P_c$.

## 3.4 Lifetime measurement

Here we present lifetime measurements of a single NV-center positioned on a 20nm $SiO_2$ spacer film on a silver substrate. Lifetime measurements are performed before and after fabrication of the HSQ width-gradient circular nanoridges around the NV-center. The NV-center is excited with a 532nm pulsed laser with a pulse width of ~100ps and a 400ns pulse period. The fluorescence photons are detected with an avalanche photo diode ($\tau - SPAD$, Pico quant). The time delay between laser pulse sync signal and photon detection is recorded by an electronic timing box (Picharp-300; Pico quant). The instrument response function of the overall system is measured to ~800ps. The measured lifetime curves exhibit a multi-exponential decay well described by the fitting a two-exponential decay curve.

$$I(t) = A_1 e^{-\frac{t}{\tau_1}} + A_2 e^{-\frac{t}{\tau_2}} + brg \tag{14}$$

We note while multi-exponential decay is typically associated with a signal arising from multiple emitters, it is here confirmed to arise from a single NV-center as $g^{(2)}(0) = 0.17$ (0.27) are measured without (with) width-gradient circular nanoridges. We speculate the two exponential decay to arise from emission from both neutral and negative charge state, clearly visible in spectrum (see Fig. 4i in main text). Fitting lifetime curves (Fig. S10) we find lifetimes $\tau_1 = 13.1 \pm 0.2$ ns ($8.9 \pm 0.2$ ns); $\tau_2 = 33.3 \pm 0.4$ ns ($27.3 \pm 0.4$ ns) without (with) width-gradient circular nanoridges, corresponding to lifetime changes of 1.5 and 1.2 respectively.

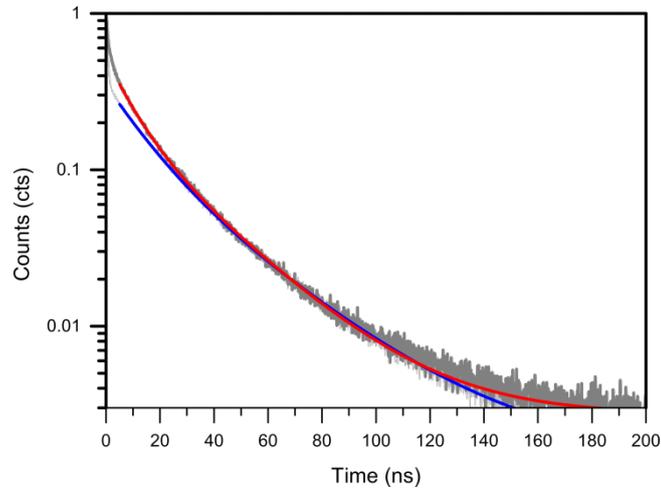

Fig. S10 Lifetime measurement of single NV-center. Experimental data is given in gray and exponential fit in blue and red, for measurement without and with width-gradient circular nanoridges respectively.